\documentclass[journal]{IEEEtran}
\usepackage{graphicx,amssymb,amsmath}
\usepackage{multicol}
\usepackage[noadjust]{cite}
\usepackage{setspace}
\usepackage{stfloats}
\usepackage{cite}
\usepackage{amsthm}
\usepackage{flushend}
\usepackage{cases,subeqnarray}
\usepackage{bm,multirow,bigstrut}
\usepackage{textcomp}
\usepackage{latexsym,bm}
\usepackage{xcolor}
\usepackage{mathtools}
\usepackage{dsfont}
\usepackage{extarrows}
\usepackage{algorithm}
\usepackage{algpseudocode}
\usepackage{amsmath}
\usepackage{amsthm}
\usepackage{booktabs}
\usepackage{caption}
\usepackage{subfigure}

\theoremstyle{plain}

\theoremstyle{plain}

\IEEEoverridecommandlockouts

\begin{document}

\title{Decentralized Coordinated Precoding Design in Cell-Free Massive MIMO Systems for URLLC}
\author{Enyu Shi, Jing Zhang, Jiayi~Zhang,~\IEEEmembership{Senior Member,~IEEE},
Derrick~Wing~Kwan Ng,~\IEEEmembership{Fellow,~IEEE,}
\\and Bo Ai,~\IEEEmembership{Fellow,~IEEE}

\thanks{This work was supported in part by the Fundamental Research Funds for the Central Universities 2022JBZX030 and 2022JBQY004, in part by Frontiers Science Center for Smart High-speed Railway System, in part by Natural Science Foundation of Jiangsu Province, Major Project under Grant BK20212002, and in part by ZTE Corporation, and State Key Laboratory of Mobile Network and Mobile Multimedia Technology. D. W. K. Ng is supported by the Australian Research Council's Discovery Project (DP210102169). (\textit{Corresponding author: Jiayi Zhang}).}
\thanks{E. Shi, J. Zhang and J. Zhang are with the School of Electronic and Information Engineering, Beijing Jiaotong University, Beijing, China. They are also with Frontiers Science Center for Smart High-speed Railway System, Beijing Jiaotong University, Beijing, China. (email: jiayizhang@bjtu.edu.cn).}
\thanks{D.~W.~K. Ng is with School of Electrical Engineering and Telecommunications, University of New South Wales, Sydney, N.S.W., Australia (email: w.k.ng@unsw.edu.au).}
\thanks{B. Ai is with State Key Laboratory of Rail Traffic Control and Safety, Beijing Jiaotong University, Beijing 100044, China (email: boai@bjtu.edu.cn).}
}
\maketitle

\begin{abstract}
Cell-free massive multiple-input multiple-output (MIMO) is a promising network to offer huge improvement of the achievable rate compared with conventional cellular massive MIMO systems. However, the commonly adopted Shannon-type achievable rate is only valid in the long block length regime that is not applicable to the emerging short-packet communication. To realize ultra-reliable and low-latency communication (URLLC) in cell-free massive MIMO systems, we optimize the precoding vector at the access points (APs) to maximize the minimum user rate in both the centralized and decentralized fashion. The design takes into account the impact of URLLC and we propose path-following algorithms (PFA) to address the considered problem which generates a sequence of advanced feasible points and converges to at least a locally optimal solution of the design problem. Moreover, we investigate the requirement of the precoding schemes, the length of the transmission duration, the number of antennas equipped at each AP, and the size of each AP cluster on the URLLC rate. Numerical results show that the decentralized PFA precoding can achieve 80\% of the 95\%-likely URLLC rate of the centralized precoding and 89\% of the average URLLC rate with only 12\% computational complexity of the centralized precoding.
\end{abstract}



\IEEEpeerreviewmaketitle
\vspace{-0.5cm}
\section{Introduction}
Ultra-reliable and low-latency communication (URLLC) is one of the generic applications required to be covered in the fifth-generation (5G) \cite{durisi2016toward,bennis2018ultrareliable,zhang2021ris}.
As a result, it has been attracted significant interests since it enables several innovative usages, especially in industrial production, such as remote heavy industrial machines operation and factory automation \cite{simsek20165g,liu2018tractable}.
However, compared with conventional communication systems, the achievable rate under URLLC is quite different since short blocklength is adopted to shorten the latency such that the classical Shannon-sense capacity no longer holds.
Specifically, the URLLC rate is a complicated function of the transmission power, the precoding vector, the bandwidth, the transmission time, and the decoding error probability \cite{polyanskiy2010channel}.
Indeed, guaranteeing URLLC represents unique challenges to resource allocation design due to the non-convexity introduced by the finite blocklength.
In the literature, much attention has been devoted to designing effective resource allocation algorithms that support URLLC \cite{she2018joint,nasir2020resource,he2021beamforming}.
However, the systems considered in these works are all cellular networks and their performance is known to be limited by severe inter-cell interference.

Cell-free massive multiple-input multiple-output (MIMO) architecture is a new promising solution to overcome the issue discussed above \cite{zhang2021local,zhang2021improving,zheng2022cell,9737367}.
It reaps the advantages of massive MIMO and network MIMO, since massive distributed access points (APs) facilitate coherent signal transmission to serve all the users without any cell boundaries \cite{bjornson2019making,ngo2017cell,zhang2020prospective}.
However, current literature focuses on the resource allocation in cell-free massive MIMO systems for URLLC is still limited.
For example, in \cite{nasir2021cell}, the authors applied the path-following algorithm (PFA) for optimizing the power allocation with a special class of conjugate beamforming to maximize the users' minimum URLLC rate and the energy efficiency.
However, an adaptive and optimized precoding design at the APs is generally more effective that the fixed one.
Besides, in \cite{lancho2022cell}, the upper bounds of the uplink and downlink decoding error probabilities (DEPs) were derived by using the saddlepoint method to support URLLC.
While the closed-form expression of DEP can characterize the performance, it is generally intractable for the the design of cooperatively efficient resource allocation.
As such, there is an emerging need for designing the precoding with the performance metric of the URLLC rate.

Motivated by the above discussion, the PFA-based precoding design for maximizing the users' minimum URLLC rate is studied in this correspondence.
First, a PFA-based centralized precoding design is proposed which generates a sequence of feasible points and converges to a locally optimal solution of the design optimization problem.
Second, we propose a decentralized PFA-based precoding design by dividing the APs into several non-overlapping cooperative clusters in which the APs only share the data and instantaneous channel state information (CSI) in each cluster to design the precoding vectors to reduce the computational complexity. Simulation results show that compared with the centralized precoding, the decentralized PFA precoding can achieve 80\% of the 95\%-likely URLLC rate and 89\% of the average URLLC rate with only 12\% of the computational complexity of the counterpart. We also investigate the impact of the precoding schemes, the length of transmission duration, and the size of the AP cluster on the URLLC rate via extensive simulations.
\section{System Model}
We consider a cell-free massive MIMO system, which consists of $L$ APs and $K$ single-antenna users that are distributed arbitrarily over a large area. We assume that each AP is equipped with $N$ antennas. Moreover, all the APs are connected with each other and a central processing unit (CPU) via dedicated fronthaul links with sufficient capacity. All APs serve all users on the same time-frequency resource through time division duplex (TDD) operation \cite{9743355}.

The channel coefficient between AP $l$ and user $k$, ${{\bf{h}}_{kl}} \in {{\mathbb{C}}^{N \times 1}}$, is assumed to follow a correlated Rayleigh fading distribution.
We adopt a classic block fading model for modeling the channels such that ${{\bf{h}}_{kl}}$ remains constant in $t$ channel uses of the time-frequency blocks and experience an independent realization in every block. Note that the channel coefficients can be acquired at the APs by existing channel estimation algorithms \cite{bjornson2017massive} and this is beyond the scope of this work as we aim to optimize the precoding for URLLC. Therefore, we assume that perfect CSI is available at the APs.

In the downlink payload data transmission phase, the received signal at user $k$ can be expressed as
${y_k} =\sum\limits_{l = 1}^L {\bf{h}}_{kl}^H{{\bf{w}}_{kl}}{s_k} + \sum\limits_{l = 1}^L {{\bf{h}}_{kl}^H} \sum\limits_{i \ne k}^K {{\bf{w}}_{il}}{s_i} + {n_k}$,
where ${s_i} \sim {{\cal N}_{\mathbb{C}}}\left( {0,1} \right)$ at AP $l$, ${{\bf{w}}_{il}} \in {{\mathbb{C}}^{N \times 1}}$ is the precoding vector for user $i$ at AP $l$, and ${n_k} \sim {{\cal N}_{\mathbb{C}}}\left( {0,{\sigma ^2}} \right)$ represents the thermal noise at user $k$.
Then, the corresponding effective signal-to-interference-plus-noise ratio (SINR) is given as
\begin{equation}
{\varphi _k} = \frac{{{{\left| {{\bf{ h}}_k^H{{\bf{w}}_k}} \right|}^2}}}{{\sum\limits_{i \ne k}^K {{{\left| {{\bf{ h}}_k^H{{\bf{w}}_i}} \right|}^2}}  + {\sigma ^2}}},
\end{equation}
where ${{\bf{h}}_k} = {\left[ {{\bf{h}}_{k1}^H, \cdots ,{\bf{h}}_{kL}^H} \right]^H} \in {{\mathbb{C}}^{LN \times 1}}$ and ${{\bf{w}}_i} = {\left[ {{\bf{w}}_{i1}^H, \cdots ,{\bf{w}}_{iL}^H} \right]^H} \in {{\mathbb{C}}^{LN \times 1}}$. By treating the inter-user interference ${\bf{h}}_{kl}^H\sum\limits_{i \ne k}^K {{{\bf{w}}_{il}}{s_i}}$ as Gaussian noise, where $p_{il}^{{\rm{dl}}} \buildrel \Delta \over =  {\left\| {{{\bf{w}}_{il}}} \right\|^2}$ is the power allocated to user $i$ at AP $l$, the achievable rate in nats/sec/Hz for user $k$ for the case of sufficiently long blocklength is given by the Shannon rate function
${{\tilde R}_k}= \ln \left( {1 + {\varphi _k}} \right)$,
and the achievable URLLC rate in nats/sec/Hz for user $k$ can be approximated as \cite[eq. (30)]{nasir2020resource}
\begin{equation}\label{URLLC Rate}
{R_k} = \ln \left( {1 + {\varphi _k}} \right) - \sqrt {\frac{1}{{tB}} \times {V_k}}  \times {Q^{ - 1}}\left( {\epsilon} \right),
\end{equation}
where $t$ is the transmission duration, $B$ is the communication bandwidth, ${V_k}$ is the channel dispersion \cite{nasir2020resource} which can be expressed as
${V_k} = 1 - \frac{1}{{{{\left( {1 + {\varphi _k}} \right)}^2}}}$,
${Q^{ - 1}}\left(  \cdot  \right)$ is the inverse of the Gaussian Q-function, i.e., $Q\left( x \right) = \int_x^\infty  {\frac{1}{{\sqrt {2\pi } }}\exp \left( { - {t^2}/2} \right)} dt$, and ${\epsilon}$ is the decoding error probability. Note that (\ref{URLLC Rate}) is the normal approximation when the channel ${{\bf{h}}_k}$ is assumed to be quasi-static and deterministic over the transmission duration $t$. The subtrahend in (\ref{URLLC Rate}) captures the rate penalty due to the finite block length, $tB$.

\section{Max-min Rate Based Precoding Design}\label{Design}
\subsection{Centralized Precoding Design}
In the centralized precoding design, the optimization of the precoding vectors takes place at the CPU, where the estimate of the global instantaneous CSI ${{{\bf{h}}}_{kl}},\forall k \in \left\{ {1, \cdots ,K} \right\},\forall l \in \left\{ {1, \cdots ,L} \right\}$, available.

The centralized max-min URLLC rate optimization problem can be expressed as
\begin{align}
&\mathop {\max }\limits_{\bf{w}} \mathop {\min }\limits_{k = 1, \cdots ,K} \left\{ {{R_k}\left( {\bf{w}} \right)} \right\}\label{P1}\tag{3a}\\
&{\rm{s.}}{\rm{t.}}\;\;\;\;\;\;\sum\limits_{k = 1}^K {{{\left\| {{{\bf{w}}_{kl}}} \right\|}^2}}  \le {p_{\max }},\forall l,\label{3b}\tag{3b}
\end{align}
where ${\bf{w}} = \left\{ {{{\bf{w}}_{kl}}:k = 1, \cdots ,K,l = 1, \cdots ,L} \right\}$ and ${p_{\max }}$ is the maximum power at each AP. The problem (\ref{P1}) is non-convex due to the URLLC rate function ${R_k}\left( {\bf{w}} \right)$. With the help of \cite{nasir2020resource}, we apply the PFA to develop a concave lower bound for ${R_k}\left( {\bf{w}} \right)$.

Without loss of generality, the URLLC rate expression for user $k$ can be rewritten as
${R_k}\left( {\bf{w}} \right) = {f_k}\left( {\bf{w}} \right) - a{g_k}\left( {\bf{w}} \right)$,
where $a = {Q^{ - 1}}\left( {\epsilon } \right)/\sqrt {t{ B}}$, ${f_k}\left( {\bf{w}} \right) = \ln \left( {1 + {\varphi _k}\left( {\bf{w}} \right)} \right)$, and ${g_k}\left( {\bf{w}} \right) = \sqrt {1 - 1/{{\left( {1 + {\varphi _k}\left( {\bf{w}} \right)} \right)}^2}}$. Now, we aim to establish a convex lower bound for ${f_k}\left( {\bf{w}} \right)$ and a concave upper bound for ${g_k}\left( {\bf{w}} \right)$.

Let ${{\bf{w}}^{\left( n \right)}}$ be a feasible point for (\ref{P1}) that is computed from the $\left( {n - 1} \right)$th iteration of the iterative PFA.

\subsubsection{Lower bounding for ${f_k}\left( {\bf{w}} \right)$}
According to \cite{nasir2020resource}, the following inequality holds for all ${\bf{x}} \in {{\mathbb{C}}^{{M_1}}},{\bf{y}} \in {{\mathbb{C}}^{{M_2}}}$ and ${\bf{\bar x}} \in {{\mathbb{C}}^{{M_1}}},{\bf{\bar y}} \in {{\mathbb{C}}^{{M_2}}}$
\begin{align}\label{I1}\tag{4}
\ln\! \left( \!\!{1\!\! +\! \frac{{{{\left\| {\bf{x}} \right\|}^2}}}{{{{\left\| {\bf{y}} \right\|}^2} \!\!+\! {\sigma ^2}}}}\!\! \right) \!\!\ge\!\! a\!\! -\! \frac{{{{\left\| {{\bf{\bar x}}} \right\|}^2}}}{{2{\cal R}\!\!\left\{ {{{{\bf{\bar x}}}^H}{\bf{x}}} \right\} \!\!- \! {{\left\| {{\bf{\bar x}}} \right\|}^2}}}\!-\! b{\left\| {\bf{x}} \right\|^2} \!\!-\! c{\left\| {\bf{y}} \right\|^2}.
\end{align}
Applying the inequality in (\ref{I1}) for $x = {\bf{h}}_k^H{{\bf{w}}_k}$, $y = {{\cal L}_k}\left( {\bf{w}} \right)$, $\bar x = {\bf{h}}_k^H{\bf{w}}_k^{\left( n \right)}$, $\bar y = {{\cal L}_k}\left( {{{\bf{w}}^{\left( n \right)}}} \right)$, where ${{\cal L}_k}\left( {\bf{w}} \right)$ arranges ${\bf{h}}_k^H{{\bf{w}}_i},i \ne k$ into a vector of dimension $K-1$, we obtain
\begin{align}\label{f_n}
{f_k}\left( {\bf{w}} \right) &\ge \bar a_k^{\left( n \right)} - \frac{{{{\left| {{\bf{h}}_k^H{\bf{w}}_k^{\left( n \right)}} \right|}^2}}}{{2{\cal R}\left\{ {{{\left( {{\bf{w}}_k^{\left( n \right)}} \right)}^H}{{{\bf{h}}}_k}{\bf{h}}_k^H{{\bf{w}}_k}} \right\} - {{\left| {{\bf{h}}_k^H{\bf{w}}_k^{\left( n \right)}} \right|}^2}}}\notag\\
&- \!\bar b_k^{\left( n \right)}{\left| {{\bf{h}}_k^H{{\bf{w}}_k}} \right|^2} \!-\! \bar c_k^{\left( n \right)}\sum\limits_{i \ne k} {{{\left| {{\bf{h}}_k^H{{\bf{w}}_i}} \right|}^2}} \!\buildrel \Delta \over = \!f_k^{\left( n \right)}\left( {\bf{w}} \right)\tag{5},
\end{align}
with the constraint of
\begin{equation}\label{trust region}\tag{6}
2{\cal R}\left\{ {{{\left( {{\bf{w}}_k^{\left( n \right)}} \right)}^H}{{{\bf{h}}}_k}{\bf{h}}_k^H{{\bf{w}}_k}} \right\} - {\left| {{\bf{h}}_k^H{\bf{w}}_k^{\left( n \right)}} \right|^2} > 0,
\end{equation}
where
$\bar a_k^{\left( n \right)} \!=\! {f_k}\left( {{{\bf{w}}^{\left( n \right)}}} \right) \!+\! 2 \!-\! \frac{{{{\left| {{\bf{h}}_k^H{\bf{w}}_k^{\left( n \right)}} \right|}^2}}}{{\beta _k^{\left( n \right)}}}\frac{{\sigma ^2}}{{\alpha _k^{\left( n \right)}}}$,
$0 \!< \!\bar b_k^{\left( n \right)} \!=\! \frac{{\bar a_k^{\left( n \right)}}}{{\beta _k^{\left( n \right)}{{\left| {{\bf{h}}_k^H{\bf{w}}_k^{\left( n \right)}} \right|}^2}}}$,
$0 \!<\! \bar c_k^{\left( n \right)} \!= \!\frac{{{{\left| {{\bf{h}}_k^H{\bf{w}}_k^{\left( n \right)}} \right|}^2}}}{{\beta _k^{\left( n \right)}\alpha _k^{\left( n \right)}}}$,
$\alpha _k^{\left( n \right)} \!\buildrel \Delta \over =\! \sum\limits_{i \ne k} \!{{{\left| {{\bf{h}}_k^H{\bf{w}}_i^{\left( n \right)}} \right|}^2}}  \!+\! {{\sigma ^2}}$,
and $\beta _k^{\left( n \right)} \!\buildrel \Delta \over = \! \sum\limits_{i = 1}^K  {{{\left| {{\bf{h}}_k^H{\bf{w}}_i^{\left( n \right)}} \right|}^2}}  \!+\! {{\sigma ^2}}$.
According to \cite{nasir2020resource}, the function $f_k^{\left( n \right)}\left( {\bf{w}} \right)$ is concave over the trust region (\ref{trust region}) and achieves the same value as  ${f_k}\left( {\bf{w}} \right)$ at ${{{\bf{w}}^{\left( n \right)}}}$,
$f_k^{\left( n \right)}\left( {{{\bf{w}}^{\left( n \right)}}} \right) = {f_k}\left( {{{\bf{w}}^{\left( n \right)}}} \right)$.

\subsubsection{Upper bounding for ${g_k}\left( {\bf{w}} \right)$}
Since the function $f\left( x \right) = \sqrt x$ is concave on $x > 0$, the following inequality for all $x > 0$ and $\bar x > 0$ holds true
\begin{align}\label{I2}\tag{7}
\sqrt x  \!=\! f\left( x \right)\le f\left( {\bar x} \right) \!+\! {\left. {\frac{{\partial \!f\!\left( x \right)}}{{\partial x}}} \right|_{x = \bar x}}\left( {x \!-\! \bar x} \right)\!=\! \frac{{\sqrt {\bar x} }}{2} \!+ \!\frac{x}{{2\sqrt {\bar x} }},
\end{align}
where $\frac{{\partial f\left( x \right)}}{{\partial x}}$ refers to the partial derivative of the function $f\left( x \right)\le f\left( {\bar x} \right)$ with respect to $x$.
Applying the inequality in (\ref{I2}) for $x = 1 - 1/{\left( {1 + {\varphi _k}\left( {\bf{w}} \right)} \right)^2}$ and $\bar x = 1 - 1/{\left( {1 + {\varphi _k}\left( {{{\bf{w}}^{\left( n \right)}}} \right)} \right)^2}$ and
using
\begin{align}\label{I3}
&{{{{\left( {\sum\limits_{i \ne k} {{{\left| {{\bf{h}}_k^H{{\bf{w}}_i}} \right|}^2}}  + {{\sigma ^2}}} \right)}^2}}}/{{{{\left( {\sum\limits_{i = 1}^K {{{\left| {{\bf{h}}_k^H{{\bf{w}}_i}} \right|}^2}}  + {{\sigma ^2}}} \right)}^2}}}\notag\\
&\ge\!\!\! \frac{{4\alpha _k^{\left( n \right)}}}{{{{\left( {\beta _k^{\left( n \right)}} \right)}^2}}}\!\!\left( \!{\sum\limits_{i \ne k}\! {\left(\! {2{\cal R}\!\left\{ \!{{{\left( \! {{\bf{h}}_k^H{\bf{w}}_i^{\left( n \right)}} \!\right)}^*}{\bf{h}}_k^H{{\bf{w}}_i}} \right\} \!\!-\! {{\left| {{\bf{h}}_k^H{\bf{w}}_i^{\left( n \right)}} \!\right|}^2}} \!\right)} } { + {{\sigma ^2}}} \!\!\right)\notag\\
&-\!\! \frac{{2{{\left( \!{\alpha _k^{\left( n \right)}} \!\right)}^2}}}{{{{\left(\! {\beta _k^{\left( n \right)}} \!\right)}^3}}}\!\!\left( \!{\sum\limits_{i = 1}^K \!{{{\left| {{\bf{h}}_k^H{{\bf{w}}_i}} \right|}^2}} \!\! +\! {{\sigma ^2}}} \!\!\right)\!-\! \frac{{{{\left( \!{\sum\limits_{i \ne k}\! {{{\left| {{\bf{h}}_k^H{{\bf{w}}_i}} \right|}^2}}  \!\!+\! {{\sigma ^2}}} \!\!\right)}^2}}}{{{{\left( {\beta _k^{\left( n \right)}} \right)}^2}}}\tag{8},
\end{align}
with the constraints of
\begin{align}
&\sum\limits_{i = 1}^K {{{\left| {{\bf{h}}_k^H{{\bf{w}}_i}} \right|}^2}}  + {{\sigma ^2}} \le 2\beta _k^{\left( n \right)},\label{cons_g1}\tag{9}\\
&\frac{1}{{{{\left( {\beta _k^{\left( n \right)}} \right)}^2}}}\left( {\sum\limits_{i = 1}^K {{{\left| {{\bf{h}}_k^H{{\bf{w}}_i}} \right|}^2}}  + {{\sigma ^2}}} \right)\le \!\! \frac{2}{{\alpha _k^{\left( n \right)}}}\notag\\
&\;{\times}\!\!\left( \! {\sum\limits_{i \ne k}\! {\left( {2{\cal R}\!\left\{ {{{\left( {{\bf{h}}_k^H{\bf{w}}_i^{\left( n \right)}} \! \right)}^*}{\bf{h}}_k^H{{\bf{w}}_i}} \!\right\}} \right.} } {\left. { \!\!-\! {{\left| {{\bf{h}}_k^H{\bf{w}}_i^{\left( n \right)}} \right|}^2}} \right) \!\!+\! \!{{\sigma ^2}}} \right),\label{cons_g2}\tag{10}
\end{align}
we have
\begin{align}\label{g_n}
{g_k}\left( {\bf{w}} \right) &\le d_k^{\left( n \right)} - \frac{{4\alpha _k^{\left( n \right)}e_k^{\left( n \right)}}}{{{{\left( {\beta _k^{\left( n \right)}} \right)}^2}}}\left( {\sum\limits_{i \ne k} {\left( {2{\cal R}\left\{ {{{\left( {{\bf{h}}_k^H{\bf{w}}_i^{\left( n \right)}} \right)}^*}{\bf{h}}_k^H{{\bf{w}}_i}} \right\}} \right.} } \right. \notag\\
&\left.{\left. { - \!{{\left| {{\bf{h}}_k^H{\bf{w}}_i^{\left( n \right)}} \right|}^2}} \right) \!\!+\! {{\sigma ^2}}} \!\right)\!\! + \!\! \frac{{2{{\left( \!{\alpha _k^{\left( n \right)}} \!\right)}^2}e_k^{\left( n \right)}}}{{{{\left( {\beta _k^{\left( n \right)}} \right)}^3}}}\!\!\left( \!{\sum\limits_{i = 1}^K \!{{{\left| {{\bf{h}}_k^H{{\bf{w}}_i}} \right|}^2}} \!\! +\!\! {{\sigma ^2}}} \!\! \right)\notag\\
&+\frac{{{{\left( {\sum\limits_{i \ne k} {{{\left| {{\bf{h}}_k^H{{\bf{w}}_i}} \right|}^2}}  + {{\sigma ^2}}} \right)}^2}e_k^{\left( n \right)}}}{{{{\left( {\beta _k^{\left( n \right)}} \right)}^2}}}\buildrel \Delta \over = g_k^{\left( n \right)}\left( {\bf{w}} \right)\tag{11},
\end{align}
where
$0 \!< \!d_k^{\left( n \right)} \!= \!\frac{{\sqrt {1 \!-\! 1/{{\left( {1 \!+\! {\varphi _k}\left( {{{\bf{w}}^{\left( n \right)}}} \right)} \right)}^2}} }}{2} \!+\! \frac{1}{{2\sqrt {1 \!-\! 1/{{\left( {1 \!+\! {\varphi _k}\left( {{{\bf{w}}^{\left( n \right)}}} \right)} \right)}^2}} }}$, and
$0 < e_k^{\left( n \right)} = \frac{1}{{2\sqrt {1 - 1/{{\left( {1 + {\varphi _k}\left( {{{\bf{w}}^{\left( n \right)}}} \right)} \right)}^2}} }}$.
The function $g_k^{\left( n \right)}\left( {\bf{w}} \right)$ is convex and achieves the same value as  ${g_k}\left( {\bf{w}} \right)$ at ${{\bf{w}}^{\left( n \right)}}$,
$g_k^{\left( n \right)}\left( {{{\bf{w}}^{\left( n \right)}}} \right) = {g_k}\left( {{{\bf{w}}^{\left( n \right)}}} \right)$.

\subsubsection{Concave Lower bound for ${R_k}\left( {\bf{w}} \right)$}
By applying (\ref{f_n}) and (\ref{g_n}), we have
${R_k}\left( {\bf{w}} \right) \ge f_k^{\left( n \right)}\left( {\bf{w}} \right) - ag_k^{\left( n \right)}\left( {\bf{w}} \right) \buildrel \Delta \over = R_k^{\left( n \right)}\left( {\bf{w}} \right)$,
under the trust region constrained by (\ref{trust region}), (\ref{cons_g1}), and (\ref{cons_g2}). The function $R_k^{\left( n \right)}\left( {\bf{w}} \right)$ is concave and matches with the function ${R_k}\left( {\bf{w}} \right)$ at ${{\bf{w}}^{\left( n \right)}}$:
\begin{equation}\label{R_n}\tag{12}
{R_k}\left( {{{\bf{w}}^{\left( n \right)}}} \right) = R_k^{\left( n \right)}\left( {{{\bf{w}}^{\left( n \right)}}} \right).
\end{equation}
At the $n$th iteration, we solve the following convex problem with the computational complexity ${\cal O}\left( {{{\left( {LNK} \right)}^3}\left( {2K + 1} \right)} \right)$ to generate the next feasible point ${\bf{w}}^{\left( {n+1} \right)}$:
\begin{equation}\label{P2}\tag{13}
\mathop {\max }\limits_{\bf{w}} \mathop {\min }\limits_{k = 1, \cdots ,K} \left\{ {R_k^{\left( n \right)}\left( {\bf{w}} \right)} \right\}
\;\;\;{\rm{s.}}{\rm{t.}}\;\;\;\text{(\ref{3b}),\;(\ref{trust region}),\;(\ref{cons_g1}),\;(\ref{cons_g2})}.
\end{equation}
According to (\ref{f_n}) and (\ref{g_n}), we can conclude that
$\mathop {\min }\limits_{k = 1, \cdots ,K} {R_k}\left( {{{\bf{w}}^{\left( {n + 1} \right)}}} \right) \ge \mathop {\min }\limits_{k = 1, \cdots ,K} {R_k}\left( {{{\bf{w}}^{\left( n \right)}}} \right),\;\forall n$,
which guarantees the monotonicity in convergence.

According to \cite{nasir2020resource,nasir2021cell,xing2020matrix1,xing2020matrix2}, it is important to have a proper initial point ${{\bf{w}}^{\left( 0 \right)}}$ with the positive URLLC rate. Thus, we start from any random point ${{\bf{w}}^{\left( 0 \right)}}$ satisfying the convex power constraint $\sum\limits_{k = 1}^K {{{\left| {{{\bf{w}}_{kl}}} \right|}^2}}  \le K,\forall l$ and (\ref{trust region}), and then iterate
\begin{equation}\label{Shannon}\tag{14}
\mathop {\max }\limits_{\bf{w}} \mathop {\min }\limits_{k = 1, \cdots ,K} f_k^{\left( n \right)}\left( {\bf{w}} \right)
\;\;\;{\rm{s.}}{\rm{t.}}\;\;\;\text{(\ref{3b})},
\end{equation}
The solution obtained by these iterations can be adopted as the feasible initial point ${{\bf{w}}^{\left( 0 \right)}}$. Finally, Algorithm 1 provides the pseudo-code for the applied path-following procedure.

\begin{algorithm}[t]
\caption{Path-Following Algorithm for Solving Problem  (\ref{P1})}
\begin{algorithmic}[1]
\State \textbf{Initialization}: Iterate the convex problem (\ref{Shannon}) until the convergence to obtain an initial point ${{\bf{w}}^{\left( 0 \right)}}$. Set $n=0$.
\State Using (\ref{f_n}) to obtain a concave lower bound for ${f_k}\left( {\bf{w}} \right)$ with constraint (\ref{trust region}).
\State Using (\ref{g_n}) to obtain a convex upper bound for ${g_k}\left( {\bf{w}} \right)$ with constraints (\ref{cons_g1}) and (\ref{cons_g2}).
\State Using (\ref{R_n}) to obtain a concave lower bound for ${R_k}\left( {\bf{w}} \right)$ under the trust region constrained by (\ref{trust region}), (\ref{cons_g1}), and (\ref{cons_g2}).
\State \textbf{Repeat until (\ref{P1}) converges} : Solve the convex problem (\ref{P2}) to generate ${\bf{w}}^{\left( {n+1} \right)}$.
\end{algorithmic}
\end{algorithm}

\subsection{Decentralized Precoding Design}
The previously proposed centralized precoding design requires all the APs to upload the instantaneous CSI to the CPU, which put a significant burden on the fronthaul signaling. Besides, the computational complexity of the centralized precoding design can be exceedingly high for a huge number of antennas. As such, there is a desire for designing the precoding in a decentralized manner which only requires local instantaneous CSI at the APs. In practice, the APs can be divided into several non-overlapping cooperation clusters in which the APs in the same cluster shares both the data and the instantaneous CSI to design the precoding vectors. The APs in different clusters only have the knowledge of the statistical CSI, such as the mean and the variance.
Note that although APs are divided into clusters, each user is served by all the APs instead of the APs in the cluster which the user resides in.

Assume each cluster contains $M$ APs, therefore, there are $L/M$ clusters in the network. As stated before, each AP can obtain the instantaneous CSI of the APs in the same cluster and the statistical CSI of the APs in different clusters. Therefore, the virtual SINR of user $k$ in cluster ${\cal{L}}$ for designing the precoding vector can be expressed as
\begin{equation}\label{VSINR-1}\tag{15}
\varphi _{k{\cal L}}^{\rm{V}}\left( {{{\bf{w}}_{k{\cal L}}}} \right)\! \! = \!\!\frac{{{{\left| {\sum\limits_{l \in {\cal L}} {{\bf{h}}_{kl}^H{{\bf{w}}_{kl}}}  + \sum\limits_{\bar l \notin {\cal L}} {{\mathbb{E}}\left\{ {{\bf{h}}_{k\bar l}^H} \right\}{{\bf{w}}_{k\bar l}}} } \right|}^2}}}{{\sum\limits_{i \ne k}^K {{{\left| {\sum\limits_{l \in {\cal L}} {{\bf{h}}_{kl}^H{{\bf{w}}_{il}}}  + \sum\limits_{\bar l \notin {\cal L}} {{\mathbb{E}}\left\{ {{\bf{h}}_{k\bar l}^H} \right\}{{\bf{w}}_{i\bar l}}} } \right|}^2}}  \!+ \!{\sigma ^2}}}.
\end{equation}
Since we consider Rayleigh fading channels, we have ${\mathbb{E}}\left\{ {{\bf{h}}_{k\bar l}^H}\right\} = {\bf{0}}$. Therefore, (\ref{VSINR-1}) can be written as 
\begin{equation}\label{VSINR-2}\tag{16}
\varphi _{k{\cal L}}^{\rm{V}}\left( {{{\bf{w}}_{k{\cal L}}}} \right) = \frac{{{{\left| {\sum\limits_{l \in {\cal L}} {{\bf{h}}_{kl}^H{{\bf{w}}_{kl}}} } \right|}^2}}}{{\sum\limits_{i \ne k}^K {{{\left| {\sum\limits_{l \in {\cal L}} {{\bf{h}}_{kl}^H{{\bf{w}}_{il}}} } \right|}^2}}  + {{\sigma ^2}}}}.
\end{equation}
The decentralized max-min URLLC rate optimization problem can be expressed as
\begin{align}\label{P_distributed}
&\mathop {\max }\limits_{{\bf{w}}_{\cal L}^{\rm{V}}} \mathop {\min }\limits_{k = 1, \cdots ,K} R_{k{\cal L}}^{\rm{V}}\left( {{\bf{w}}_{\cal L}^{\rm{V}}} \right)\notag\\
&\;{\rm{s.}}{\rm{t.}}\;\;\;\;\;\sum\limits_{k = 1}^K {{{\left| {{{\bf{w}}_{k{\cal L}}}} \right|}^2}}  \le {p_{\max}},\forall l \in {\cal L},\tag{17}
\end{align}
where ${\bf{w}}_{\cal L}^{\rm{V}}$ represents the precoding vectors designed for all the users by APs in cluster ${\cal{L}}$ according to (\ref{VSINR-2}), and
$R_{k{\cal L}}^{\rm{V}}\left( {{\bf{w}}_{k{\cal L}}^{\rm{V}}} \right) = \ln \!\left( {1 + \varphi _{k{\cal L}}^{\rm{V}}\left( {{\bf{w}}_{k{\cal L}}^{\rm{V}}} \right)} \right) - \sqrt {\frac{1}{{tB}} \times V_{k{\cal L}}^{\rm{V}}}  \times{Q^{ - 1}}\left( \epsilon  \right)$,
$V_{k{\cal L}}^{\rm{V}} = 1 - \frac{1}{{{{\left( {1 + \varphi _{k{\cal L}}^{\rm{V}}\left( {{\bf{w}}_{k{\cal L}}^{\rm{V}}} \right)} \right)}^2}}}$.

The problem (\ref{P_distributed}) can be solved in a similar approach as the one for (\ref{P1}). When the problem (\ref{P_distributed}) has been solved for all the clusters, we can obtain the precoding vector for user $k$ by
\begin{equation}\label{w}\tag{18}
{{\bf{w}}_k} = {\left[ {{{\left( {{\bf{w}}_{k1}^{\rm{V}}} \right)}^H}, \cdots ,{{\left( {{\bf{w}}_{k\left( {L/M} \right)}^{\rm{V}}} \right)}^H}} \right]^H}.
\end{equation}
Then, the URLLC rate of user $k$ can be obtained by computing (\ref{URLLC Rate}) using the precoding vector obtained from (\ref{w}). The computational complexity for each iteration in decentralized precoding design is ${\cal O}\left( {{{\left( {\left( {\frac{L}{M}} \right)NK} \right)}^3}\left( {2K + 1} \right)} \right)$. Compared with the centralized precoding, the computational complexity decreased by $M^3$.

\section{Numerical Results}
In this section, we evaluate the performance of the proposed PFA precoding design for the centralized and the decentralized fashion and investigate the impact of the precoding schemes, the length of transmission duration $t$, the number of antennas equipped at each AP $N$, and the size of the AP cluster $M$ on the URLLC rate. We first describe our adopted simulation parameters.
We adopt the similar parameters setting as in \cite{ngo2017cell} as the basis to establish our simulation system model.
$L$ APs and $K$ users are deployed in a rectangular area of $96\times48$ $\text{m}^{2}$. In particular, the APs are deployed on a rectangle grid. The area is wrapped around at the edges to avoid the boundary effects \cite{ngo2017cell}. The horizontal spacing between APs are $24$ m, and the vertical spacing is $12$ m. The $K$ users are deployed randomly. We adopt a similar propagation model as in \cite{bjornson2019making}. Besides, we set $L = 16$, $\tau_p = 3$, and $\epsilon = 10^{-5}$. Note that in all the figures, the achievable rates are calculated in bits/s/Hz.

\begin{figure}[t!]
\centering
\includegraphics[width=3in]{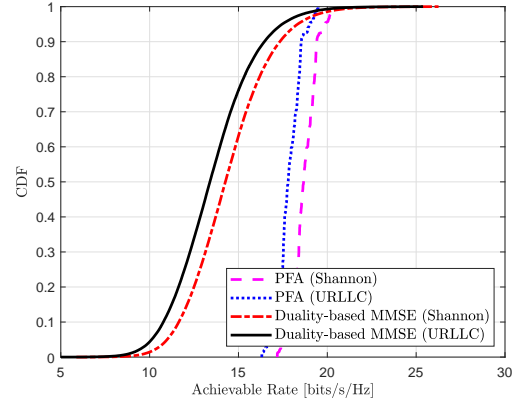}
\caption{CDF of the achievable rate achieved by the centralized PFA precoding and the duality-based MMSE precoding with $t = 0.05$ ms, $B = 1$ MHz, $K = 6$, and $N = 4$.}
\label{fig_centralized_MMSE_vs_PFA}
\end{figure}

Fig. \ref{fig_centralized_MMSE_vs_PFA} shows the cumulative distribution functions (CDFs) of the achievable rate per user achieved by the proposed PFA centralized precoding and the duality-based MMSE precoding with $t = 0.05$ ms, $B = 1$ MHz, $K = 6$, and $N = 4$ which is given by
\begin{equation}\label{MMSE}\tag{19}
{{\bf{w}}_k} = \frac{{{{\bf{v}}_k}}}{\left\| {{{\bf{v}}_{kl}}} \right\|},\;\;\;
{{\bf{v}}_k} = p{\left( {\sum\limits_{i = 1}^K p {{{\bf{h}}}_i}{\bf{h}}_i^H  + {\sigma ^2}{{\bf{I}}_{LN}}} \right)^{ - 1}}{{{\bf{h}}}_k},
\end{equation}
where $p$ is the transmit power intend for each user at each AP. It can be observed that the proposed PFA centralized precoding scheme performs very well. The achievable rate per user distribution with the proposed PFA centralized precoding almost uniformly outperforms the duality-based MMSE precoding, and the former is more steeper.
Specifically, applying the PFA centralized precoding leads to 32\% improvement in terms of average URLLC rate and 65\% improvement in terms of 95\%-likely URLLC rate. Note that the duality-based MMSE precoding in (\ref{MMSE}) is only a heuristic solution utilizing the uplink-downlink duality and cannot effectively minimize the MSE ${\mathbb{E}}\left\{ {\left. {{{\left| {{y_k} - {s_k}} \right|}^2}} \right|{{{\bf{h}}}_{kl}}} \right\}$.
Moreover, compared with the PFA centralized precoding, the duality-based MMSE precoding has a lower computational complexity since it only requires $\frac{{{N^2}{L^2}K + NLK}}{2} + \frac{{{N^3}{L^3} - NL}}{3} + {N^2}{L^2}$ complex-valued multiplications. Besides, as expected, the performance of Shannon rate serves as a performance upper bound of the URLLC rate at the expense of infinitely long code length.

\begin{figure}[t!]
\centering
\includegraphics[width=3in]{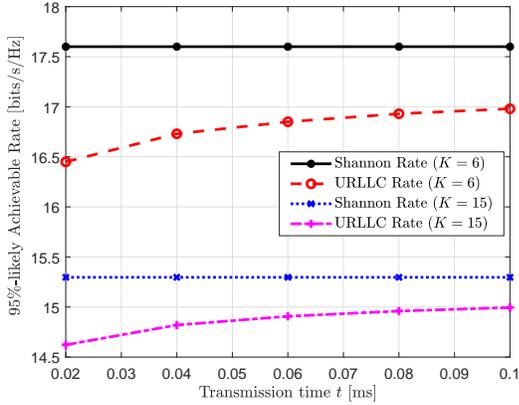}
\caption{Optimized 95\%-likely achievable rate versus the transmission time $t$ with $N = 4$ and $B = 1$ MHz.}
\label{fig_T}
\end{figure}

Fig. \ref{fig_T} plots the optimized 95\%-likely achievable rate by Algorithm 1 versus the transmission time $t$ with $N = 4$ and $B = 1$ MHz .
As expected, the URLLC rate increases along with the transmission time $t$ according to the expression of the URLLC rate.
Note that the Shannon rate is fixed since it is computed assuming a sufficient long blocklength, e.g., $t \to \infty$.
Besides, when the number of user increases from 6 to 15, we can observe that the achievable rate decreases since there are more users competing for limited resources that reduces the flexibility of the resource allocation for effective beamforming.
The performance gap between the Shannon rate and URLLC rate is also reduced with the increasing number of users as the performance of these two scheme is limited by the user with the weakest channel gain.

\begin{figure}[t!]
\centering
\includegraphics[width=3in]{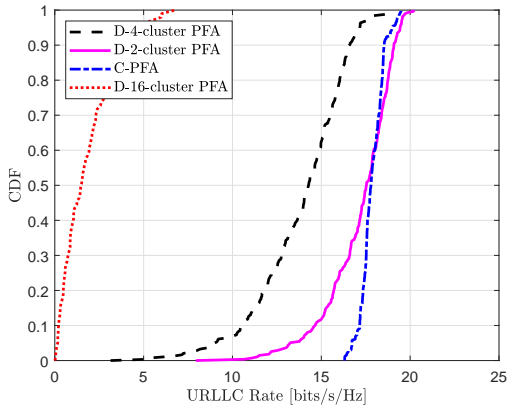}
\caption{CDF of the URLLC rate achieved by the PFA precoding in the centralized and decentralized way with $t = 0.05$ ms and $N = 4$.}
\label{fig_URLLC}
\end{figure}


Fig. \ref{fig_URLLC} shows the performance of the PFA precoding in the centralized and decentralized fashion in terms of the URLLC rate.
The curve ``C-PFA'' represents the URLLC rate computed using the centralized PFA precoding design. Also, the curve ``D-4-cluster'', ``D-2-cluster'', and ``D-16-cluster'' stand for the performance of the decentralized PFA precoding design with 4 APs, 8 APs, and 1 AP in each cluster, respectively.
The first observation from Fig. \ref{fig_URLLC} is that compared with the centralized PFA precoding, the 95\%-likely URLLC rate with the decentralized PFA precoding is generally lower.
This is because when the decentralized PFA precoding is adopted, only the instantaneous CSI within the cluster and the statistical CSI outside the cluster are used for optimization in each cluster.
As there is a mismatch between the statistical CSI and the instantaneous CSI, the optimization for the decentralized setting is less effective for the utilization of the system resources.
Besides, the performance of the 2-cluster decentralized PFA precoding outperforms the centralized PFA precoding for the strong users. The reason is that the performance of the centralized PFA precoding is always limited by the worst-case users, since substantial resources are allocated to equalize all the SINRs, while the decentralized PFA precoding benefits from being more scalable.
Compared with the 2-cluster decentralized PFA precoding, when adopting the 4-cluster or 16-cluster decentralized PFA precoding, the mismatch between the statistical CSI and the instantaneous CSI is pronounced, so the performance is the worse.
Specifically, compared with the centralized precoding, the 95\%-likely URLLC rate is reduced from 16.73 bits/s/Hz to 13.25 bits/s/Hz with the 2-cluster decentralized PFA precoding and to 8.95 bits/s/Hz with the 4-cluster decentralized PFA precoding.
Moreover, when the fully distributed 16-cluster decentralized PFA precoding is adopted, the 95\%-likely URLLC rate is only 0.17 bits/s/Hz.
However, since the computational complexity is also reduced, the performance loss of adopting the 2-cluster decentralized PFA precoding instead of the centralized precoding is tolerable.
In particular, the 2-cluster decentralized PFA precoding achieves 80\% of the 95\%-likely URLLC rate, 89\% of the average URLLC rate, and 12\% of the computational complexity of the centralized precoding.
The second observation is that the CDF of users' URLLC rate is not as steep as the counterpart when the decentralized PFA precoding design is adopted. The reason is that the optimization target of each cluster contains virtual SINR rather than the actual SINR, leading to under utilisation of system resources.


\section{Conclusion}
In this correspondence, we considered the precoding design in the cell-free massive MIMO system for URLLC in the centralized and decentralized fashion. PFA was designed for maximizing the users' minimum URLLC rate and its performance was evaluated with different settings of the transmission time, the number of antennas per AP, and the size of the AP cluster. Simulation results showed that the centralized PFA precoding design can effectively improve the performance of 95\%-likely achievable rate and the decentralized PFA precoding with a reasonable setting can approach the performance of the former but with low computational complexity. In the future, we will jointly optimize the precoding vector, the cluster formation, and the number of APs in each cluster in a distributed fashion for URLLC.

\bibliographystyle{IEEEtran}
\bibliography{IEEEabrv,Ref_URLLC}

\begin{thebibliography}{10}
\providecommand{\url}[1]{#1}
\csname url@samestyle\endcsname
\providecommand{\newblock}{\relax}
\providecommand{\bibinfo}[2]{#2}
\providecommand{\BIBentrySTDinterwordspacing}{\spaceskip=0pt\relax}
\providecommand{\BIBentryALTinterwordstretchfactor}{4}
\providecommand{\BIBentryALTinterwordspacing}{\spaceskip=\fontdimen2\font plus
\BIBentryALTinterwordstretchfactor\fontdimen3\font minus
  \fontdimen4\font\relax}
\providecommand{\BIBforeignlanguage}[2]{{%
\expandafter\ifx\csname l@#1\endcsname\relax
\typeout{** WARNING: IEEEtran.bst: No hyphenation pattern has been}%
\typeout{** loaded for the language `#1'. Using the pattern for}%
\typeout{** the default language instead.}%
\else
\language=\csname l@#1\endcsname
\fi
#2}}
\providecommand{\BIBdecl}{\relax}
\BIBdecl

\bibitem{durisi2016toward}
G.~Durisi, T.~Koch, and P.~Popovski, ``Toward massive, ultrareliable, and
  low-latency wireless communication with short packets,'' \emph{Proceeding of
  IEEE}, vol. 104, no.~9, pp. 1711--1726, 2016.

\bibitem{bennis2018ultrareliable}
M.~Bennis, M.~Debbah, and H.~V. Poor, ``Ultrareliable and low-latency wireless
  communication: Tail, risk, and scale,'' \emph{Proceeding of IEEE}, vol. 106,
  no.~10, pp. 1834--1853, 2018.

\bibitem{zhang2021ris}
J.~Zhang, H.~Liu, Q.~Wu, Y.~Jin, Y.~Chen, B.~Ai, S.~Jin, and T.~J. Cui,
  ``{RIS}-aided next-generation high-speed train communications: {C}hallenges,
  solutions, and future directions,'' \emph{IEEE Wireless Commun.}, vol.~28,
  no.~6, pp. 145--151, Dec. 2021.

\bibitem{simsek20165g}
M.~Simsek, A.~Aijaz, M.~Dohler, J.~Sachs, and G.~Fettweis, ``{5G}-enabled
  tactile internet,'' \emph{IEEE J. Sel. Areas Commun.}, vol.~34, no.~3, pp.
  460--473, Mar. 2016.

\bibitem{liu2018tractable}
L.~Liu, Y.~Zhou, W.~Zhuang, J.~Yuan, and L.~Tian, ``Tractable coverage analysis
  for hexagonal macrocell-based heterogeneous {UDNs} with adaptive
  interference-aware {CoMP},'' \emph{IEEE Trans. Wireless Commun.}, vol.~18,
  no.~1, pp. 503--517, Jan. 2018.

\bibitem{polyanskiy2010channel}
Y.~Polyanskiy, H.~V. Poor, and S.~Verd{\'u}, ``Channel coding rate in the
  finite blocklength regime,'' \emph{IEEE Trans. Inf. Theory}, vol.~56, no.~5,
  pp. 2307--2359, May 2010.

\bibitem{she2018joint}
C.~She, C.~Yang, and T.~Q. Quek, ``Joint uplink and downlink resource
  configuration for ultra-reliable and low-latency communications,'' \emph{IEEE
  Tran. Commun.}, vol.~66, no.~5, pp. 2266--2280, May 2018.

\bibitem{nasir2020resource}
A.~A. Nasir, H.~D. Tuan, H.~H. Nguyen, M.~Debbah, and H.~V. Poor, ``Resource
  allocation and beamforming design in the short blocklength regime for
  {URLLC},'' \emph{IEEE Trans. Wireless Commun.}, vol.~20, no.~2, pp.
  1321--1335, Feb. 2020.

\bibitem{he2021beamforming}
S.~He, Z.~An, J.~Zhu, J.~Zhang, Y.~Huang, and Y.~Zhang, ``Beamforming design
  for multiuser {URLLC} with finite blocklength transmission,'' \emph{IEEE
  Transa. Wireless Commun.}, vol.~20, no.~12, pp. 8096--8109, Dec. 2021.

\bibitem{zhang2021local}
J.~Zhang, J.~Zhang, E.~Bj{\"o}rnson, and B.~Ai, ``Local partial zero-forcing
  combining for cell-free massive {MIMO} systems,'' \emph{IEEE Trans. Commun.},
  vol.~69, no.~12, pp. 8459--8473, Dec. 2021.

\bibitem{zhang2021improving}
J.~Zhang, J.~Zhang, D.~W.~K. Ng, S.~Jin, and B.~Ai, ``Improving sum-rate of
  cell-free massive {MIMO} with expanded compute-and-forward,'' \emph{IEEE
  Trans. Signal Process.}, vol.~70, no.~11, pp. 202--215, Nov. 2021.

\bibitem{zheng2022cell}
J.~Zheng, J.~Zhang, E.~Bj{\"o}rnson, Z.~Li, and B.~Ai, ``Cell-free massive
  {MIMO-OFDM} for high-speed train communications,'' \emph{IEEE J. Sel. Areas
  Commun., to appear}, 2022.

\bibitem{9737367}
Z.~Wang, J.~Zhang, B.~Ai, C.~Yuen, and M.~Debbah, ``Uplink performance of
  cell-free massive {MIMO} with multi-antenna users over jointly-correlated
  {R}ayleigh fading channels,'' \emph{IEEE Trans. Wireless Commun., to appear}.

\bibitem{bjornson2019making}
E.~Bj{\"o}rnson and L.~Sanguinetti, ``Making cell-free massive {MIMO}
  competitive with {MMSE} processing and centralized implementation,''
  \emph{IEEE Trans. Wireless Commun.}, vol.~19, no.~1, pp. 77--90, Jan. 2019.

\bibitem{ngo2017cell}
H.~Q. Ngo, A.~Ashikhmin, H.~Yang, E.~G. Larsson, and T.~L. Marzetta,
  ``Cell-free massive {MIMO} versus small cells,'' \emph{IEEE Trans. Wireless
  Commun.}, vol.~16, no.~3, pp. 1834--1850, Mar. 2017.

\bibitem{zhang2020prospective}
J.~Zhang, E.~Bj{\"o}rnson, M.~Matthaiou, D.~W.~K. Ng, H.~Yang, and D.~J. Love,
  ``Prospective multiple antenna technologies for beyond {5G},'' \emph{IEEE J.
  Sel. Areas in Commun.}, vol.~38, no.~8, pp. 1637--1660, Aug. 2020.

\bibitem{nasir2021cell}
A.~A. Nasir, H.~D. Tuan, H.~Q. Ngo, T.~Q. Duong, and H.~V. Poor, ``Cell-free
  massive {MIMO} in the short blocklength regime for {URLLC},'' \emph{IEEE
  Trans. Wireless Commun.}, vol.~20, no.~9, pp. 5861--5871, Sep. 2021.

\bibitem{lancho2022cell}
A.~Lancho, G.~Durisi, and L.~Sanguinetti, ``Cell-free massive {MIMO} for
  {URLLC}: {A} finite-blocklength analysis,'' \emph{arXiv:2207.00856}, 2022.

\bibitem{9743355}
E.~Shi, J.~Zhang, S.~Chen, J.~Zheng, Y.~Zhang, D.~W. Kwan~Ng, and B.~Ai,
  ``Wireless energy transfer in {RIS}-aided cell-free massive {MIMO} systems:
  Opportunities and challenges,'' \emph{IEEE Commun. Mag.}, vol.~60, no.~3, pp.
  26--32, Mar. 2022.

\bibitem{bjornson2017massive}
E.~Bj{\"o}rnson, J.~Hoydis, and L.~Sanguinetti, ``Massive {MIMO} networks:
  Spectral, energy, and hardware efficiency,'' \emph{Found. Trends Signal
  Process.}, vol.~11, no. 3-4, pp. 154--655, 2017.

\bibitem{xing2020matrix1}
C.~Xing, S.~Wang, S.~Chen, S.~Ma, H.~V. Poor, and L.~Hanzo, ``Matrix-monotonic
  optimization $-$ part {I}: {S}ingle-variable optimization,'' \emph{IEEE
  Trans. Signal Process.}, vol.~69, no.~11, pp. 738--754, Nov. 2020.

\bibitem{xing2020matrix2}
------, ``Matrix-monotonic optimization $-$ part {II}: {M}ulti-variable
  optimization,'' \emph{IEEE Transactions on Signal Processing}, vol.~69,
  no.~11, pp. 179--194, Nov. 2020.

\end{thebibliography}

\end{document}